\documentclass[prb,twocolumn,showpacs]{revtex4}
\usepackage{graphicx}

\catcode`\@=11
\def\simle{\mathrel{\mathpalette\@versim<}}   
\def\simge{\mathrel{\mathpalette\@versim>}}   
\def\@versim#1#2{\lower2.5pt\vbox{\baselineskip0pt \lineskip-.5pt
   \ialign{$\m@th#1\hfil##\hfil$\crcr#2\crcr\sim\crcr}}}
\newcommand{\mib}[1]{\mbox{\boldmath $#1$}} 

\begin{document}

\title{
Monte Carlo Study of Doping Change and Disorder Effect \\
on Double Exchange Ferromagnetism}

\author{
Yukitoshi Motome
}

\affiliation{
RIKEN (The Institute of Physical and Chemical Research), 
2-1 Hirosawa, Wako, Saitama 351-0198, Japan 
}

\author{
Nobuo Furukawa
}

\affiliation{
Department of Physics, Aoyama Gakuin University, 
5-10-1 Fuchinobe, Sagamihara, Kanagawa 229-8558, Japan
}

\date{\today}

\begin{abstract}
Phase diagram and critical properties are studied 
for three-dimensional double exchange model with and without quenched disorder. 
Employing the Monte Carlo method and the systematic analysis 
on the finite-size effect, we estimate the Curie temperature and 
the critical exponent as functions of the doping concentration and 
the strength of the random potential. 
The Curie temperature well scales to the kinetic energy of electrons 
in the ground state as expected for this kinetics-driven ferromagnetism. 
The universality class of this transition is described 
by the short-range Heisenberg fixed point. 
The results are compared with the experimental results 
in the colossal magnetoresistance manganites.
\end{abstract}



\maketitle

\section{Introduction}
\label{Sec:Introduction}

Since Zener's pioneering work, 
\cite{Zener1951}
the double exchange (DE) model and its extensions have been studied 
to understand the variety of magnetic and transport properties  
including the colossal magnetoresistance (CMR) in perovskite manganese oxides. 
\cite{Anderson1955,deGennes1960,Furukawa1999a,Dagotto2001}
The original DE model, which contains the single-band electrons 
interacting with the localized spins through the Hund's-rule coupling 
(the Hamiltonian will be explicitly given in Sec.~\ref{Sec:Model}), 
qualitatively explains the stability of 
the ferromagnetic metallic state and the negative magnetoresistance. 
There, the parallel configuration of localized spins leads to 
the kinetic energy gain of electrons, and vice versa. 
The effective ferromagnetic interaction between localized spins 
is mediated by the motion of electrons, and is called the DE interaction. 
Despite of the long survey for more than half century, however, 
the quantitative aspects of the thermodynamics of this model 
have not been fully explored thus far. 
This is mainly because the system is in the strongly-correlated regime 
in the sense that the Hund's-rule coupling is much larger than 
the bandwidth of electrons. 
Both thermal and spatial fluctuations are crucial 
in the thermodynamics of this itinerant electron system, 
and they are difficult to handle in a controllable manner. 

Monte Carlo (MC) calculation is one of the powerful tools to treat 
such strong correlations properly. 
In the present problem, since the wavefunction of itinerant electrons 
is extended, 
it is crucial to make a systematic finite-size scaling analysis 
for obtaining a reliable result in the thermodynamic limit.
Recently, there have been some developments in the MC algorithm
which enable us to handle larger size systems than ever 
in a reasonable computational time. 
\cite{Motome1999,FurukawaPREPRINT,Alonso2001}
The importance of spatial fluctuations and finite-size effects has been 
examined by comparing the MC results with the mean-field results or
the dynamical mean-field approximation (DMFA) results. 
\cite{Motome2000}
The reliable estimate of the Curie temperature $T_{\rm C}$ 
has been obtained by the systematic scaling analysis. 
\cite{Motome2000,Alonso2001,MotomePREPRINTa}
The critical exponents have been also estimated, 
and it is shown that the exponents are consistent with 
those of the Heisenberg spin model with short-range interactions. 
\cite{Motome2000,Alonso2001,MotomePREPRINTa,Motome2001,Furukawa2002}

These previous MC calculations have been mainly performed 
at the doping concentration $x=0.5$ (0.5 electrons per site on average), 
where the kinetic energy of electrons is maximum, namely, 
$T_{\rm C}$ becomes the highest. 
In real compounds, for instance, in La$_{1-x}$Sr$_x$MnO$_3$, 
the ferromagnetic metallic phase is stabilized 
at $0.15 \simle x \simle 0.6$, and 
$T_{\rm C}$ becomes maximum at $x \simeq 0.3$ and 
slightly reduces for $x > 0.3$. 
\cite{Urushibara1995}
This might be due to an instability
toward the A-type antiferromagnetic state 
or the charge-ordered state near $x=0.5$. 
\cite{Goodenough1955,Kuwahara1997,Akimoto1998}
These instabilities are beyond the simple DE model. 
The $x$ dependence of $T_{\rm C}$ for $0.1 \simle x \simle 0.3$ 
in La$_{1-x}$Sr$_x$MnO$_3$
has been favorably compared with the DMFA results. 
\cite{Furukawa1995}
However, the recent MC study has revealed the insufficiency of DMFA 
and the importance of spatial fluctuations as mentioned above. 
One of the purposes of this work is to determine 
$T_{\rm C}$ for wide regions of $x$ 
by applying the advanced MC technique and
to clarify the phase diagram of the DE model. 
We will compare the numerical results with the experimental results quantitatively.

Another purpose of this work is to clarify 
the disorder effect on the critical properties of 
the ferromagnetic transition in the DE model. 
Disorder suppresses the kinetic energy of electrons
and reduces $T_{\rm C}$. 
Although this reduction has been also studied by DMFA 
\cite{Allub1996,Letfulov2001,Auslender2001,Narimanov2002}
and the MC calculation for small size clusters, 
\cite{Motome2002a}
here we give more precise estimates employing the advanced MC method and 
the systematic finite-size scaling. 
Through the quantitative comparison between $T_{\rm C}$ and 
the kinetic energy of electrons, we examine how the kinetics governs 
the DE ferromagnetism in the disordered case. 
We also estimate the critical exponent to clarify the universality class
in the disordered case. 
These are also motivated by the experiments 
on the {\it A}-site substitution in {\it A}MnO$_3$ at a fixed valence $x$
which indicate the relevance of the chemical disorder of the {\it A}-site ions. 
\cite{Radaelli1997,Rodriguez-Martinez1996,Coey1995,Saitoh1999} 
We will make a quantitative comparison between theory and experiment. 

This paper is organized as follows. 
In Sec.~\ref{Sec:Model and Method}, we introduce the DE model 
including the random potential. 
The MC method as well as the details of numerical conditions 
is also described here. 
In Sec.~\ref{Sec:Results}, we show the numerical results. 
$T_{\rm C}$ and the critical exponent are estimated 
by the systematic finite-size scaling 
in both cases with and without the random potential. 
We discuss the numerical results in comparison with the experimental results 
in CMR manganites in Sec.~\ref{Sec:Discussions}. 
Section \ref{Sec:Summary} is devoted to summary.

\section{Model and Method}
\label{Sec:Model and Method}

\subsection{Model}
\label{Sec:Model}

The DE model considered here consists of itinerant electrons in the single band 
which interact with localized spins at each site 
through the Hund's-rule coupling. 
The Hamiltonian is given in the form 
\cite{Zener1951}
\begin{equation}
{\cal H}_{\rm DE} = - \sum_{\langle ij\rangle \sigma} 
t \ (c_{i\sigma}^\dagger c_{j\sigma} + {\rm h.c.}) 
- J_{\rm H} \sum_i \mib{\sigma}_i \cdot \mib{S}_i, 
\label{eq:H_DE}
\end{equation}
where the first term describes the electron hopping 
between the nearest-neighbor sites and 
the second term represents the Hund's-rule coupling 
between the Pauli spin $\mib{\sigma}$ of electrons 
and the localized spin $\mib{S}$ 
(the coupling is ferromagnetic, namely, $J_{\rm H}$ is positive). 
Additionally, we take account of the the random on-site potential as
\begin{equation}
{\cal H}_\varepsilon = \sum_{i\sigma} \varepsilon_i 
c_{i\sigma}^\dagger c_{i\sigma}, 
\label{eq:H_varepsilon}
\end{equation}
where we consider the binary distribution, namely, 
$\varepsilon_i$ takes the value of $\pm \Delta$ 
with equal probability randomly in each site. 
The total Hamiltonian is given by 
${\cal H} = {\cal H}_{\rm DE} + {\cal H}_\varepsilon$.

In the following, for simplicity, we consider 
the limit of $J_{\rm H} \rightarrow \infty$. 
In this limit, the electron spin $\mib{\sigma}$ is completely 
parallel to the localized spin $\mib{S}$ in each site, and 
states with $\mib{\sigma}$ antiparallel to $\mib{S}$ are projected out. 
This simplifies the model to the effective spinless-fermion model 
in the form 
\cite{Anderson1955}
\begin{equation}
{\cal H} = - \sum_{\langle ij \rangle} \tilde{t}_{ij} 
( \tilde{c}_i^\dagger \tilde{c}_j + {\rm h.c.}) 
+ \sum_i \varepsilon_i \tilde{c}_i^\dagger \tilde{c}_i, 
\label{eq:H}
\end{equation}
where the transfer integral $\tilde{t}_{ij}$ depends on 
the relative angle of localized spins at $i$ and $j$th site as 
\begin{equation}
\tilde{t}_{ij} = t \Big(
\cos\frac{\theta_i}{2} \cos\frac{\theta_j}{2} +
\sin\frac{\theta_i}{2} \sin\frac{\theta_j}{2} 
{\rm e}^{- {\rm i} (\phi_i - \phi_j)} \Big). 
\label{eq:t_ij}
\end{equation}
Here, the angles $\theta_i$ and $\phi_i$ are defined as 
$S_i^x = S \sin\theta_i \cos\phi_i$, 
$S_i^y = S \sin\theta_i \sin\phi_i$, and
$S_i^z = S \cos\theta_i$, 
where $S = | \mib{S}_i |$ is the magnitude of the localized spin. 
Thus, the transfer integral becomes a complex variable 
whose amplitude is proportional to $\cos(\theta_i - \theta_j)/2$ and 
phase is governed by the so-called Berry phase 
$\exp\{-{\rm i} (\phi_i - \phi_j)\}$. 

One more simplification we introduce here is to treat the localized spins 
in the classical limit of $S \rightarrow \infty$. 
Then, the present model (\ref{eq:H}) describes 
the strong interplay between the quantum itinerant electrons and 
the classical localized spins. 
If the configuration of the localized spins $\{ \mib{S}_i \}$ is frozen, 
the problem is simply the free electrons in the random magnetic field. 
However, in the present problem, the localized spins are 
not the fixed external magnetic field but internal degrees of freedom 
of the system. 
The configurations of $\{ \mib{S}_i \}$ are determined 
thermodynamically through the interaction with itinerant electrons. 
Namely, thermal equilibrium is achieved to optimize 
$\{ \mib{S}_i \}$ which reconciles the kinetics of electrons and 
the entropy of the localized spins.

\subsection{Method and Numerical conditions}
\label{Sec:Method}

In the following, we calculate the critical properties 
of the ferromagnetic transition in model (\ref{eq:H}) 
defined on the three-dimensional cubic lattice. 
We employ the truncated polynomial-expansion MC (TPEMC) method 
which is recently developed by the authors. 
\cite{FurukawaPREPRINT}
This technique is based on the polynomial-expansion MC method, 
\cite{Motome1999}
and has advantage in the computational cost 
by introducing effective truncations in the polynomial expansion. 
Readers are referred to Ref.~\onlinecite{FurukawaPREPRINT} 
for the details of the algorithm. 
Using this method, we calculate the physical quantities 
in the finite-size clusters of $N = L^3$ from $L=6$ to $16$. 
Systematic analyses on the finite-size corrections 
are performed for those series. 

We typically perform 10000 MC samplings for measurements 
after 1000 steps for thermalization 
in the absence of the disorder. 
The results are divided into five bins to estimate 
the statistical errors by the variance among the bins. 
In the presence of the disorder, 
for a given configuration of $\{ \varepsilon_i \}$, 
we typically perform 1000 MC samplings for measurements 
after 1000 steps for thermalization. 
We repeat this for typically 16 different configurations
to estimate the errors by taking the random average among the results 
for each $\{ \varepsilon_i \}$.  
We take the half bandwidth $W = 6t$ at $J_{\rm H} = \Delta = 0$ 
as an energy unit. 

We choose twisted boundary conditions 
to reduce the ground state degeneracy in the finite-size systems 
\cite{Assaad2002}
by introducing the magnetic flux $\Phi = \pi/4$, $\pi/2$, and $3\pi/4$ 
in the $x$, $y$, and $z$ direction, respectively. 
The magnetic flux $\Phi$ is included by the so-called Peierls factor 
in the phase of the transfer integrals as 
$\tilde{t}_{ij} \rightarrow \tilde{t}_{ij} \exp({\rm i} \Phi / L)$. 
This enables us to reach the converged results 
in smaller order of the polynomial expansion. 
We confirmed that, for the range of the parameters in the present work, 
the polynomial expansion up to the 8th order is enough 
to obtain the `exact' results in the limit of the infinite order. 

In the TPEMC calculations, we take the threshold values 
in the truncations as 
$10^{-4}$ for the matrix product and
$10^{-5}$ for the trace operation. 
We confirm that the truncations with these thresholds do not 
change the results beyond the errors. 

In the following, we calculate the magnetization 
at a fixed doping concentration $x$ as a function of temperature. 
($x \equiv 1 - \sum_i \langle \tilde{c}_i^\dagger \tilde{c}_i \rangle / N$, 
where the bracket denotes the thermal average for the grand canonical ensemble.) 
For that purpose, we need to control the chemical potential at each temperature 
because the band structure changes as magnetic correlations develop in this system. 
\cite{Furukawa1997}
(The case of $x=0.5$ is special because $x$ is fixed 
for $\mu=0$ due to the particle-hole symmetry.)
Instead of this laborious procedure, we fix the chemical potential $\mu$ 
so that the target value of $x$ is realized at $T \sim T_{\rm C}$. 
This leads to $T$-dependence of $x$, however, it is small 
in the parameter range of interests, and 
does not harm the results because the magnetization is not 
so sensitive to the small change of $x$. 
We confirmed that the magnetization does not show any $x$ dependence 
beyond its errors for this small deviation of $x$. 
The values of the chemical potential are taken as 
$\mu = 0.095$, $0.205$, and $0.310$ for $x = 0.4$, $0.3$, and $0.2$, respectively. 
The results are shown with the errorbars of $x$ 
which represent the small change of $x$ 
in the corresponding temperature range.

\section{Results}
\label{Sec:Results}

\subsection{Phase diagram in the absence of the disorder}
\label{Sec:pure}

\begin{figure}
\includegraphics[width=7cm]{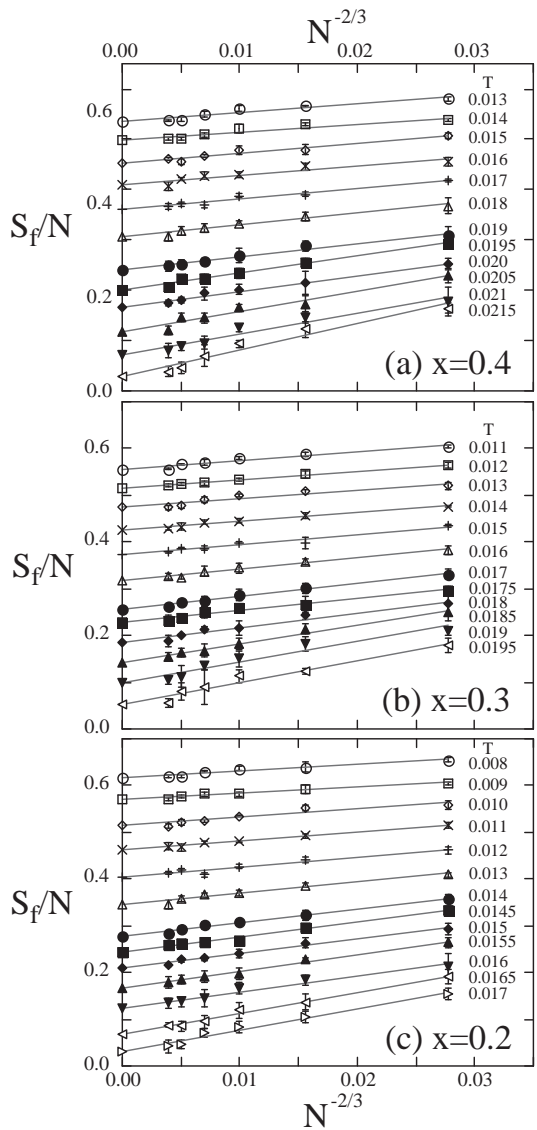}
\caption{
System-size extrapolation of the spin structure factor in the pure case 
for (a) $x=0.4$, (b) $x=0.3$, and (c) $x=0.2$. 
Symbols at $N \rightarrow \infty$ show the extrapolated values. 
} 
\label{fig:S(0) pure} 
\end{figure}

First, we study the pure case without the random potential, namely, $\Delta=0$. 
Figure~\ref{fig:S(0) pure} shows the system-size extrapolation of 
the ferromagnetic component of the spin structure factor, 
$S_{\rm f} = \sum_{ij} \langle \mib{S}_i \cdot \mib{S}_j \rangle / N$, 
divided by the system size $N$.  
All the data well scale to $N^{-2/3}$, 
which is consistent with the $\mib{k}^2$ dependence of 
the energy cutoff of magnons. 
\cite{Furukawa1996}
From the extrapolated values, 
we obtain the magnetization in the thermodynamic limit 
as $M = \lim_{N \rightarrow \infty} \sqrt{S_{\rm f}/N}$. 

\begin{figure}
\includegraphics[width=7cm]{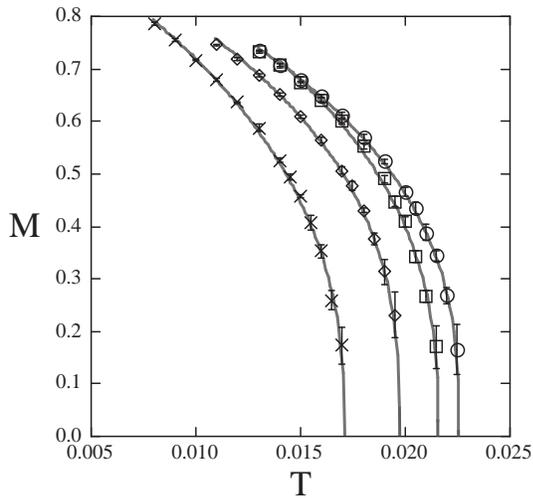}
\caption{
Temperature dependence of the magnetization in the thermodynamic limit. 
Circles, squares, diamonds, and crosses represent 
the data for $x=0.5, 0.4, 0.3$, and $0.2$, respectively. 
The data for $x=0.5$ are from Ref.~\onlinecite{MotomePREPRINTa}. 
The gray curves are the scaling fit by eq.~(\ref{eq:scaling}).
} 
\label{fig:M pure} 
\end{figure}

Figure~\ref{fig:M pure} summarizes the temperature dependence of 
the magnetization in the thermodynamic limit. 
The data for $x=0.5$ are taken from Ref.~\onlinecite{MotomePREPRINTa}. 
The results are well fitted by the scaling form 
\begin{equation}
M \propto (T_{\rm C} - T)^\beta, 
\label{eq:scaling}
\end{equation}
as shown by the gray curves in this figure. 
From these fittings, we estimate the Curie temperature $T_{\rm C}$ and
the critical exponent $\beta$ for each value of $x$. 

\begin{figure}
\includegraphics[width=7cm]{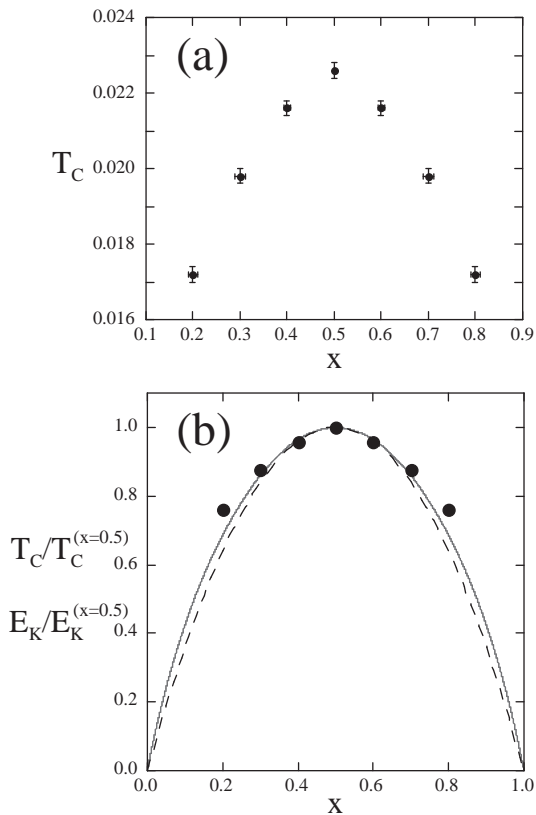}
\caption{
(a) $x$ dependence of the Curie temperature $T_{\rm C}$. 
(b) Comparison of $x$ dependences of $T_{\rm C}$ (circles), 
the kinetic energy $E_{\rm K}$ (gray curve), and $x(1-x)$ (dashed curve). 
All the data are normalized by their value at $x=0.5$ for comparison.
} 
\label{fig:Tc pure} 
\end{figure}

The estimates of $T_{\rm C}$ are summarized in Fig.~\ref{fig:Tc pure} (a). 
The results are shown also for $x>0.5$ by using 
the particle-hole symmetry, namely, $T_{\rm C}(x) = T_{\rm C}(1-x)$. 
In Fig.~\ref{fig:Tc pure} (b), 
the $x$ dependence of $T_{\rm C}$ is compared with 
the kinetic energy in the ground state (perfectly polarized state), $E_{\rm K}$,  
and the functional form of $x(1-x)$ proposed in Ref.~\onlinecite{Varma1996}. 
The data are normalized by their value at $x=0.5$ for comparison. 
As shown in the figure, 
the MC results almost scale to the kinetic energy of electrons, 
which confirms that the DE ferromagnetism is governed 
by the kinetics of electrons.
From this scaling, 
we obtain the relation $T_{\rm C}/|E_{\rm K}/N| = 0.13-0.15$. 
This enables us to estimate $T_{\rm C}$ approximately 
from the ground state quantity 
which is easily obtained in this DE system. 

The estimates of $\beta$ by the scaling fit in Fig.~\ref{fig:M pure} 
agree with the Heisenberg exponent $\beta = 0.365$ 
(Ref.~\onlinecite{LeGuillou1977})
within the errors for entire range of $x$. 
This is consistent with the previous MC study at $x=0.5$ 
which has revealed that the universality class of this ferromagnetic transition 
belongs to that of the short-range Heisenberg model. 
\cite{Motome2000,MotomePREPRINTa}

\subsection{Disorder effect}
\label{Sec:disorder}

\begin{figure}
\includegraphics[width=7cm]{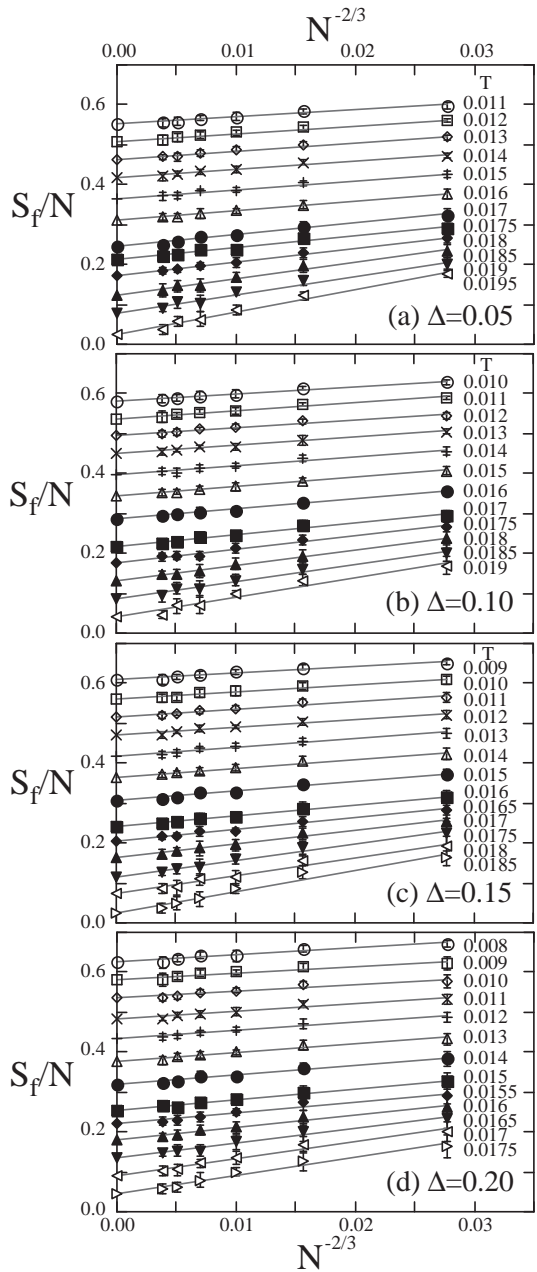}
\caption{
System-size extrapolation of the spin structure factor at $x=0.3$ 
for (a) $\Delta=0.05$, (b) $\Delta=0.1$, (c) $\Delta=0.15$, and (d) $\Delta=0.2$. 
Symbols at $N \rightarrow \infty$ show the extrapolated values. 
} 
\label{fig:S(0) random} 
\end{figure}

\begin{figure}
\includegraphics[width=7cm]{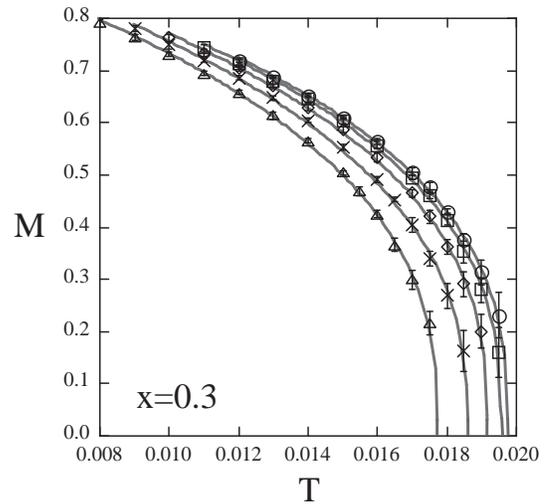}
\caption{
Temperature dependence of the magnetization in the thermodynamic limit at $x=0.3$. 
Circles, squares, diamonds, crosses, and triangles represent 
the data for $\Delta=0.0, 0.05, 0.1, 0.15$, and $0.2$, respectively. 
The gray curves are the scaling fit by eq.~(\ref{eq:scaling}).
} 
\label{fig:M random} 
\end{figure}

Next, we study the effect of the random on-site potential at $x=0.3$.
Figure~\ref{fig:S(0) random} shows the system-size extrapolation of 
the spin structure factor by varying the strength of the random potential $\Delta$. 
Even in these disordered cases, the data well scale to $N^{-2/3}$ 
as in the pure case in Fig.~\ref{fig:S(0) pure}. 
This is consistent with recent studies for the spin excitation spectrum 
which predict the $\mib{k}^2$ magnon excitation even in the presence of disorder. 
\cite{Motome2002b,Motome2003,MotomePREPRINTb}
The magnetization calculated from the extrapolated values are summarized 
in Fig.~\ref{fig:M random}. 
The random potential decreases the magnetization 
because it reduces the kinetic energy, namely, 
the DE ferromagnetic interaction. 

\begin{figure}
\includegraphics[width=7cm]{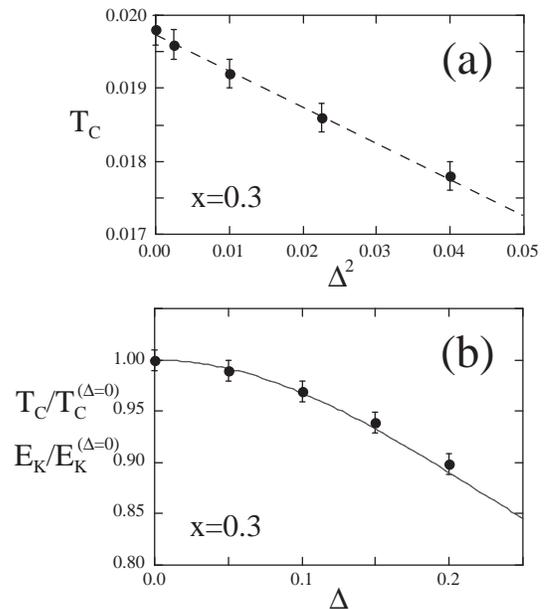}
\caption{
(a) $T_{\rm C}$ at $x=0.3$ plotted as a function of $\Delta^2$. 
The line shows the linear fit.
(b) Comparison of $\Delta$ dependences of $T_{\rm C}$ (circles) and
the kinetic energy $E_{\rm K}$ (gray curve). 
The data are normalized by their value at $\Delta=0$ for comparison.
} 
\label{fig:Tc random} 
\end{figure}

We apply the scaling of eq.~(\ref{eq:scaling}) in this disordered case also, 
and obtain the estimates of $T_{\rm C}$ in Fig.~\ref{fig:Tc random} (a). 
$T_{\rm C}$ appears to scale to $\Delta^2$ in the weak-disorder regime. 
In Fig.~\ref{fig:Tc random} (b), we compare $\Delta$ dependence of 
$T_{\rm C}$ with the kinetic energy of electrons in the ground state, and
find that $T_{\rm C}$ well scales to the kinetic energy.
This indicates that, also in the presence of the disorder,
the DE ferromagnetism is governed by the kinetics of electrons.

\begin{figure}
\includegraphics[width=7cm]{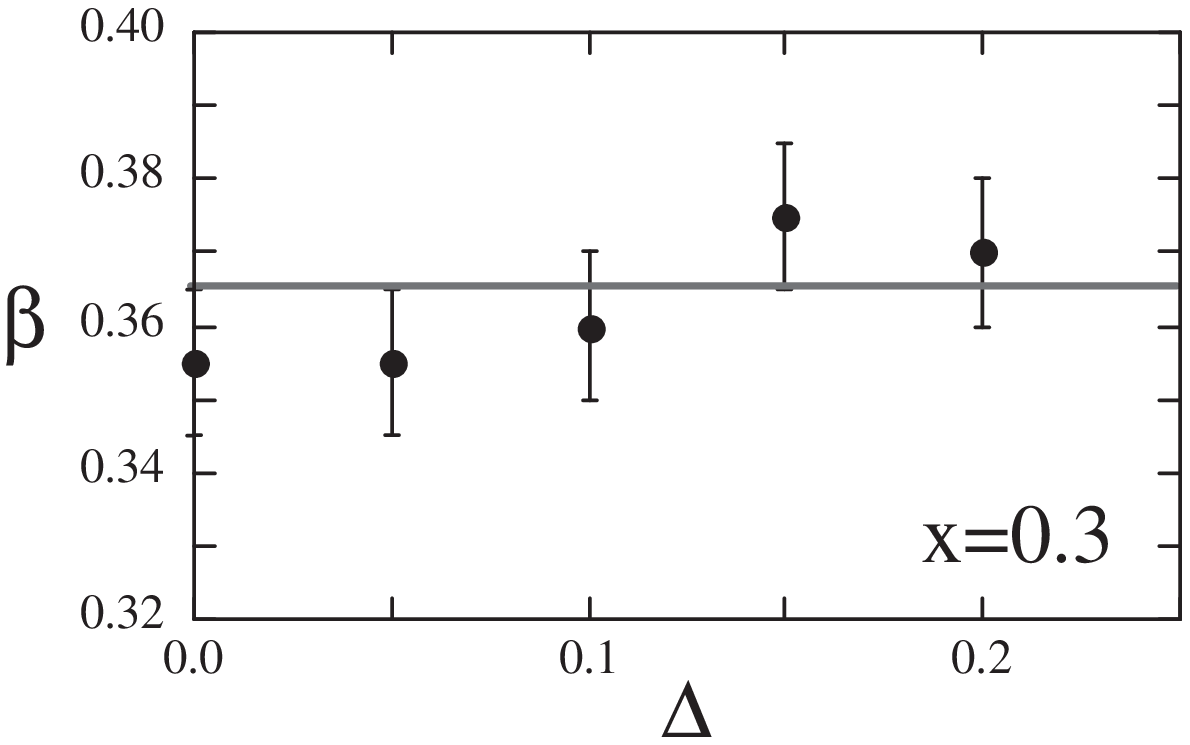}
\caption{
$\Delta$ dependence of the critical exponent $\beta$. 
The horizontal line represents the Heisenberg exponent $\beta=0.365$.
} 
\label{fig:beta random} 
\end{figure}

From the scaling fit in Fig.~\ref{fig:M random}, 
the critical exponent $\beta$ is also estimated. 
The results are shown in Fig.~\ref{fig:beta random}. 
The exponent is consistent with the Heisenberg value $\beta=0.365$ 
(Ref.~\onlinecite{LeGuillou1977}) 
even in the presence of the disorder. 
This indicates that the disorder is irrelevant and does not change 
the universality class in this transition. 

\section{Discussions}
\label{Sec:Discussions}

\subsection{Comparison with La$_{1-x}$Sr$_x$MnO$_3$}
\label{Sec:LSMO}

La$_{1-x}$Sr$_x$MnO$_3$ near $x=0.3$ has been considered to be 
a canonical DE system in which many aspects of the thermodynamics 
are successfully explained by the DE model (\ref{eq:H_DE}) alone. 
\cite{Furukawa1999a,Motome2000}
Here, we compare the estimates of $T_{\rm C}$ obtained in Sec.~\ref{Sec:pure} 
with the experimental values in this compound. 
Comparing the MC estimate $T_{\rm C}^{\rm (MC)} = 0.0198W$ 
with the experimental value $T_{\rm C}^{\rm (exp)} = 369$K 
\cite{Urushibara1995}
at $x=0.3$, 
we obtain the half bandwidth as $W \simeq 1.6$eV. 
This value is larger than the estimates by the band calculations
($W \sim 1$eV), 
\cite{Hamada1995,Pickett1997,Papaconstantopoulos1998}
however we note that this is the bare value in the case of $J_{\rm H} = 0$ 
and that a large $J_{\rm H}$ generally renormalizes the bandwidth. 

\begin{figure}
\includegraphics[width=7cm]{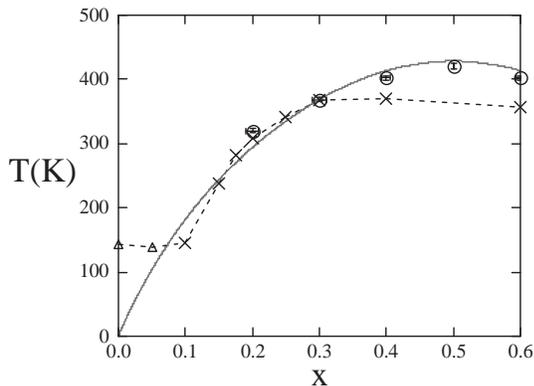}
\caption{
Comparison between the MC results of $T_{\rm C}$ (circles) and 
the experimental values in La$_{1-x}$Sr$_x$MnO$_3$ (crosses). 
The kinetic energy in the ground state is also shown (gray curve). 
All the data are normalized to agree with each other at $x=0.3$ for comparison. 
Triangles represent the antiferromagnetic transition temperature 
observed in the low $x$ regime. 
Experimental data are from Ref.~\onlinecite{Urushibara1995}.
} 
\label{fig:Tc exp} 
\end{figure}

Figure~\ref{fig:Tc exp} shows the comparison of $x$ dependence of $T_{\rm C}$ 
between theory and experiment. 
We normalize $T_{\rm C}^{\rm (MC)}$ as well as 
the kinetic energy of electrons 
to agree with $T_{\rm C}^{\rm (exp)}$ at $x=0.3$. 
We note that, in the range of $0.15 \simle x \simle 0.3$, 
$T_{\rm C}^{\rm (exp)}$ well scales to the MC results and the kinetic energy. 
This agreement has been claimed in the DMFA results. 
\cite{Furukawa1995}
For $x \simge 0.4$, $T_{\rm C}^{\rm (exp)}$ is suppressed and 
shows a deviation from this scaling. 
This might be due to the instability toward 
the A-type antiferromagnetic state or the CE-type charge-ordered state 
observed near $x \simeq 0.5$ in many CMR manganites. 
\cite{Goodenough1955,Kuwahara1997,Akimoto1998}
Our data suggest that if such instabilities are absent, 
the Curie temperature can become higher up to $\sim 420$K. 

For the critical exponents, experimental results in this compound 
are still controversial. 
The estimates are scattered from the short-range Heisenberg ones 
to the mean-field ones. 
\cite{Martin1996,Lofland1997,Vasiliu-Doloc1998a,Mohan1998,Ghosh1998,Schwartz2000}
Our results indicate that if the DE interaction plays a dominant role 
in the ferromagnetic transition, 
the Heisenberg universality class should be observed. 
\cite{Motome2000,MotomePREPRINTa,Motome2001,Furukawa2002}
Further experimental studies are desired. 

\subsection{Disorder effect in the {\it A}-site substitution}
\label{Sec:A-site}

Many experiments have indicated that the disorder is important 
in the {\it A}-site substitution in {\it A}MnO$_3$ at a fixed doping concentration. 
\cite{Radaelli1997,Rodriguez-Martinez1996,Coey1995,Saitoh1999}
There, the chemical disorder from the random distribution of the {\it A} ions 
with different ionic radii 
is considered to disturb the electronic state in Mn-O-Mn network 
electrostatically and structurally. 
For instance, $T_{\rm C}$ decreases about $30$\% 
from La$_{0.7}$Sr$_{0.3}$MnO$_3$ to La$_{0.7}$Ca$_{0.3}$MnO$_3$, 
and this cannot be explained by the pure DE model (\ref{eq:H_DE}) 
since the estimated change of the transfer integral is only about $2$\%. 
\cite{Radaelli1997}
The MC results in Sec.~\ref{Sec:disorder} shows that 
$T_{\rm C}$ decreases substantially by the disorder. 
Figure~\ref{fig:Tc random} (b) suggests that 
the $30$\% decrease of $T_{\rm C}$ might be achieved at $\Delta \sim 0.4-0.5$. 
This corresponds to $\Delta \sim 0.6-0.8$eV if we assume $W=1.6$eV 
as in Sec.~\ref{Sec:LSMO}. 
The estimate of the disorder strength appears to be consistent with 
the potential fluctuation due to the alloying effect 
predicted by the band calculations. 
\cite{Pickett1997}

The disorder effect has been also studied in the spin excitation spectrum recently. 
\cite{Motome2002b,Motome2003,MotomePREPRINTb}
The results well explain the spectral anomalies which show up 
in the {\it A}-site substituted materials. 
\cite{Hwang1998,Vasiliu-Doloc1998b,Dai2000,Biotteau2001}
This and the above observation in the decrease of $T_{\rm C}$ 
consistently indicate the importance of the disorder 
in the {\it A}-site substitution in CMR manganites. 

In the present MC study, we consider only the diagonal disorder 
of the random on-site potential. 
In real materials, the {\it A}-site disorder may affect the electronic state 
in other ways, for instance, as 
the off-diagonal disorder of the random transfer integrals. 
In the study of the spin excitation spectrum, 
it was shown that various types of disorder bring about 
the universal results in the spectrum. 
\cite{Motome2002b,Motome2003,MotomePREPRINTb}
We speculate that this is the case also in the thermodynamics, namely, 
that $T_{\rm C}$ scales to the kinetic energy and 
the critical exponents are for the Heisenberg universality class 
irrespective of the types of disorder.  

For further substitution which introduces larger difference in ionic radius, 
experimentally, the ferromagnetic state is taken over 
by the charge-ordered state concomitant with the Jahn-Teller distortion. 
\cite{Goodenough1955,Kuwahara1997,Akimoto1998}
The phase diagram shows the multicritical behavior which 
indicates a strong competition between different phases. 
\cite{Tomioka2002,AkahoshiPREPRINT,Nakajima2002}
Disorder effect on the multicritical phenomena 
has attracted much attentions recently. 
\cite{Moreo1999,Burgy2001,MotomePREPRINTc}
The emergence of the charge-ordered state suggests that, 
finally, another element such as the electron-phonon interaction 
becomes important when we approach to the multicritical point. 
However, we consider that, in the regime far from the multicritical point, 
such another element may be less important and 
the disorder plays a primary role.

\section{Summary}
\label{Sec:Summary}

We have studied the phase diagram and the universality class 
of the ferromagnetic transition in the three-dimensional double exchange model 
with and without the random potential. 
The truncated polynomial-expansion Monte Carlo method 
has been employed to calculate large size clusters 
without uncontrolled or biased approximations. 
The Curie temperature and the critical exponent have been estimated 
by applying the systematic finite-size scaling analysis 
up to $16^3$ site clusters. 
For both changes of the doping concentration and 
the strength of the random potential, 
we found that the Curie temperature $T_{\rm C}$ well scales to 
the kinetic energy of electrons per site in the ground state $E_{\rm K}/N$. 
This is a consequence of the fact that 
the kinetics of electrons governs the ferromagnetism in this system. 
From this scaling, we have obtained the approximate relation 
$T_{\rm C}/|E_{\rm K}/N| = 0.13-0.15$, 
which is useful since the ground state quantity is easily calculated 
in the double exchange system. 
In both cases with and without disorder, 
estimates of the critical exponent are consistent with 
that of the Heisenberg spin model with short-range interaction. 
This indicates that the ferromagnetic transition in the double exchange systems 
belongs to the short-range Heisenberg universality class. 
We have compared the results with the experimental results 
in Sr doping in La$_{1-x}$Sr$_x$MnO$_3$ and 
in the ionic radius control by the {\it A}-site substitution in {\it A}MnO$_3$. 

\section*{Acknowledgment}

The authors thank H. Nakata for helpful support
in developing parallel-processing systems.
The computations have been performed mainly 
using the facilities in the AOYAMA+ project
(http://www.phys.aoyama.ac.jp/\~{}aoyama+).
This work is supported by  ``a Grant-in-Aid from
the Ministry of Education, Culture, Sports, Science, and Technology''.



\end{document}